\documentclass[aps,singlecolumn,notitlepage,longbibliography, superscriptaddress]{revtex4-1}

\usepackage{graphicx}
\usepackage{dcolumn}
\usepackage{bm}
\usepackage[utf8]{inputenc}
\usepackage[T1]{fontenc}
\usepackage{mathptmx}
\usepackage{amsmath}

\begin{document}

\title{Stress-controlled frequency tuning and parametric amplification of the vibrations of coupled nanomembranes}

\author{Sepideh Naserbakht}
\author{Andreas Naesby}
\author{Aur\'{e}lien Dantan}
\email{dantan@phys.au.dk}

\affiliation{Department of Physics and Astronomy, University of Aarhus, DK-8000 Aarhus C, Denmark}

\begin{abstract}
Noninvasive tuning of the mechanical resonance frequencies of suspended parallel nanomembranes in various monolithic arrays is achieved by piezoelectric control of their tensile stress. Parametric amplification of their thermal fluctuations is shown to be enhanced by the piezoelectric actuation and amplification factors of up to 20 dB in the sub-parametric oscillation threshold regime are observed.
\end{abstract}

\date{\today}
\maketitle

\section{Introduction}

Suspended nano- and microresonators are ubiquitous in a wide range of sensing, photonics and telecommunication applications, for many of which a certain degree of tunability of their mechanical properties may be desirable~\cite{Ekinci2005,Li2007}. Dynamical tuning of the mechanics can be achieved by modifying the stress of the resonant structures, e.g. capacitively~\cite{Ekinci2005,Kozinsky2006}, electro-~\cite{Jun2006,Huttel2009,Barton2012} or photothermally~\cite{Jockel2011,Inoue2017}, by bending~\cite{Verbridge2007} or by heating~\cite{St-Gelais2019}. In general, such a stress control may allow for modifying not only the mechanical resonance frequencies and quality factors, but also the nonlinear response of the resonators~\cite{Ekinci2005,Almog2006,Antoni2012,Thomas2013}. Such a tunability is particularly interesting for \textit{arrays} consisting of multiple resonators, for instance to match or enhance their collective response and sensing capabilities~\cite{Buks2002,Sage2018}, to engineer and investigate complex collective dynamics~\cite{Shim2007,Matheny2014,Matheny2019}, or to coherently manipulate phonons between them~\cite{Bagheri2011,Mahboob2012,Faust2013,Okamoto2013,Mahboob2014,Mathew2016,Yang2019}.

Suspended drum-shaped resonators, made of low loss material such as silicon nitride and possessing high mechanical quality factors, can be efficiently coupled to electromagnetic fields, whether in the optical or microwave/radiofrequency domains~\cite{Thompson2008,Wilson2009,PurdySCI2013,Bagci2014,Andrews2014,Yuan2015,Xu2016,Rossi2018}. Having multiple such membrane resonators~\cite{Nair2017,Piergentili2018,Gartner2018,Wei2019} simultaneously interacting with cavity fields opens for a number of exciting applications, e.g., collectively enhanced optomechanics~\cite{Xuereb2012,Xuereb2013,Seok2012,Kipf2014}, optomechanical synchronization~\cite{Bemani2017}, phonon transport~\cite{Xuereb2014,Xuereb2015} or entanglement and multimode squeezing generation~\cite{Bhattacharya2008,Hartmann2008,Nielsen2017,Patil2015,Pontin2016}.

We investigate here the tuning of the linear and nonlinear mechanical properties of suspended silicon nitride square drums by the application of a piezoelectrically controlled force to the chip supporting the drums. In this scheme, recently implemented with a single~\cite{Wu2018} and pairs of membranes~\cite{Wei2019,Naserbakht2019}, the compression of the frame caused by the piezoelectric force modifies the tensile stress of the silicon nitride films, thereby allowing for reversibly tuning the mechanical resonance frequencies without deteriorating their quality factors. We report here on the simultaneous tuning of the mechanical resonance frequencies of the membranes of various monolithic double-membrane arrays, and show for instance that the modes of membranes with close resonance frequencies can be tuned to degeneracy by the application of a bias voltage to the piezoelectric element. This tunability would be essential in the abovementioned applications involving optomechanical arrays. In addition to radiation pressure, such drum resonators naturally lend themselves to fluid pressure measurements, and the tuning of their mechanical properties within such compact monolithic arrays would also be highly interesting in connection with the realization of squeeze film pressure sensors~\cite{Naesby2017,Naserbakht2019sna}.

Furthermore, as we recently reported in Ref.~\cite{Naserbakht2019}, such a piezoelectric stress control allows for enhancing the resonators' nonlinear response to dynamical actuation. We expand here on this demonstration by performing detailed studies of the parametric excitation of the thermal fluctuations of the fundamental drummodes of two closely lying membranes in a monolithic array, for which amplification factors of up to 20 dB in the sub-parametric oscillation threshold regime are observed. The observed noise spectra and gains are in excellent agreement with the predictions of a simple phase-averaged subthreshold parametric amplification model.

Mechanical parametric amplification~\cite{Rugar1991} being an ubiquitous tool in electro-opto-mechanical systems~\cite{Midolo2018} to investigate and exploit nonlinear dynamics in a wide range of MEMS applications and resonators~\cite{Turner1998,Carr2000,Ekinci2005,Mahboob2008,Karabalin2010,Antoni2012,Okamoto2013,Thomas2013,Seitner2017,Huber2019,Bothner2019}, the results demonstrated here with suspended silicon nitride drum resonators open for interesting applications involving thermomechanical squeezing~\cite{Szorkovszky2010, Farace2012,Szorkovszky2013,Mahboob2014,Patil2015,Pontin2016,Wu2018} and amplification~\cite{Lemonde2016,Levitan2016,Bothner2019} in cavity electro/optomechanics, among others.

\section{Electromechanical resonators and experimental setup}

\begin{figure}[h]
\centering\includegraphics[width=0.6\columnwidth]{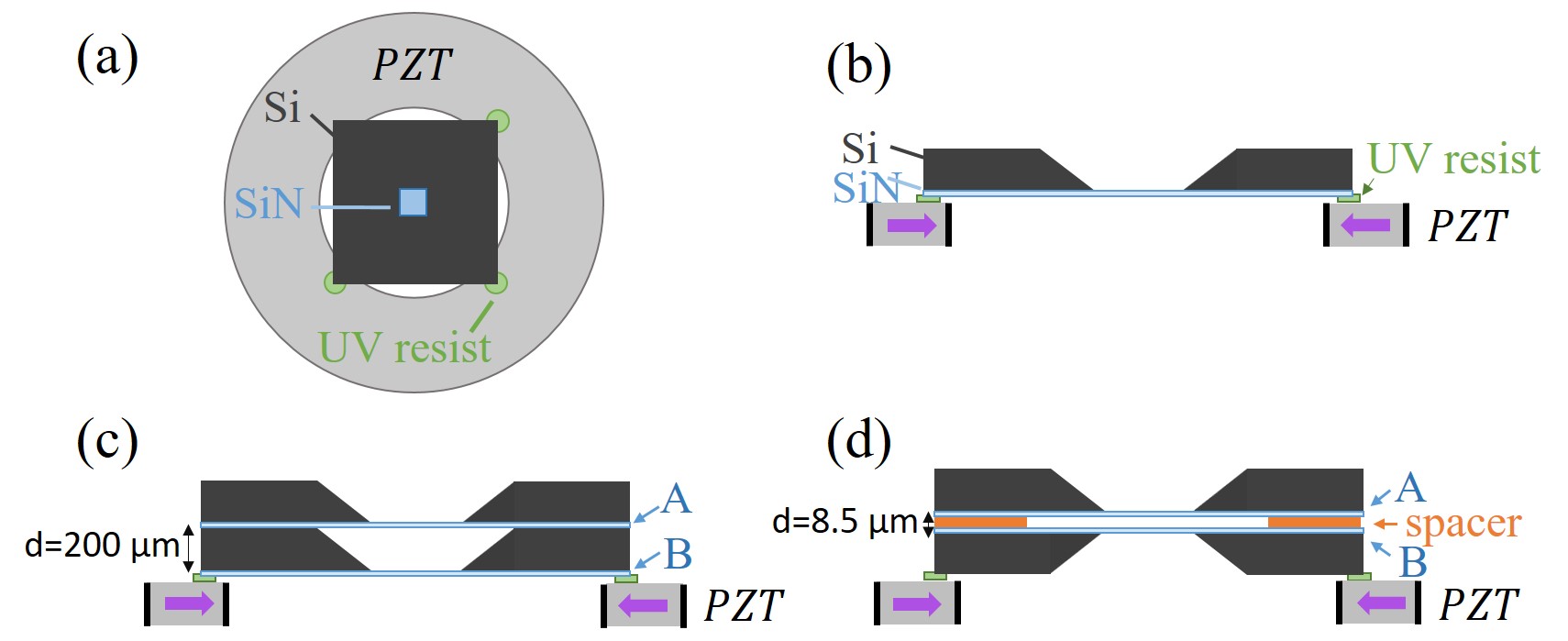}
\caption{(a) Topview schematic of the membrane chip mounted on a piezoelectric ring actuator. (b)-(d): sideview schematics of the single- or double-membrane electromechanical resonators studied. The purple arrows inside the piezoelectric transducer (PZT) indicate the direction of the compressive force for a positive voltage difference between the outer and inner electrodes.}
\label{fig:scheme}
\end{figure}

The membrane resonators used in this work are commercial~\cite{Norcada}, high tensile stress (0.7-0.9 GPa), stochiometric SiN square drums with lateral dimension 500 $\mu$m and thickness 100 nm, deposited on a Si frame with lateral dimension 5 mm and thickness of either 200 or 500 $\mu$m. As described in Ref.~\cite{Naserbakht2019} and depicted in Figs.~\ref{fig:scheme}a and b, three corners of a Si chip were glued by application and curing of a small dab of UV-resist (OrmoComp, Micro resist technology GmbH) onto a piezoelectric ring actuator with 6 mm inner diameter (Noliac NAC2123) to which both dc and ac voltages can be applied. Following the method of Ref.~\cite{Nair2017}, double-membrane arrays were made by gluing the two chips together either on top of each other (Fig.~\ref{fig:scheme}c), so that the intermembrane separation was given by the Si frame thickness (200 $\mu$m in this case), or with a spacer in between the SiN films setting the separation between the suspended membranes to be $\sim 8.5$ $\mu$m (Fig.~\ref{fig:scheme}d).  

\begin{figure}[h]
\centering\includegraphics[width=0.4\columnwidth]{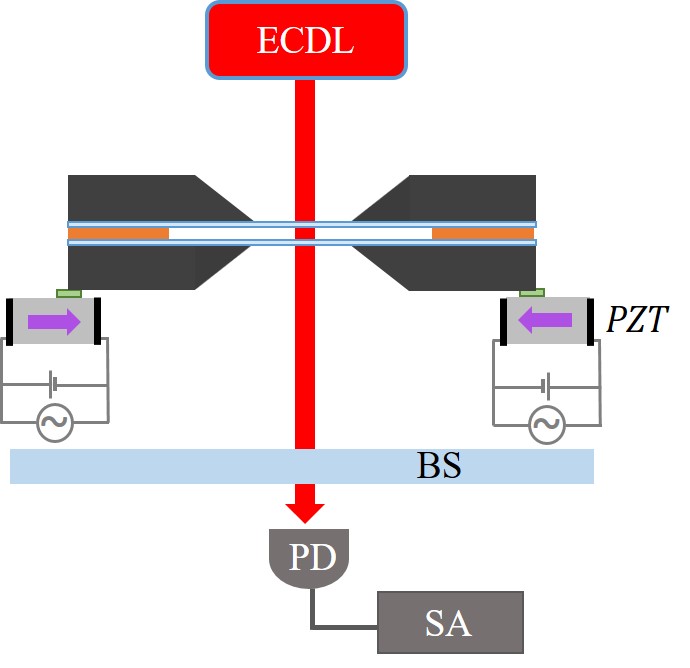}
\caption{Schematic experimental setup for the piezoelectric actuation and detection of the membrane vibrations (shown here for the array geometry of Fig.~\ref{fig:scheme}d). ECDL: external cavity diode laser, BS: beamsplitter mirror, PZT: piezoelectric transducer, PD: photodiode, SA: spectrum analyzer.}
\label{fig:setup}
\end{figure}

The vibrations of the membranes can then be measured by monitoring the transmission of monochromatic light ($\sim 900$ nm) issued from an external cavity diode laser through a linear Fabry-Perot interferometer constituted by the membrane--or membrane array--and a 50:50 beamsplitter mirror positioned parallel to the membranes at a distance of approximately 7 mm (Fig.~\ref{fig:setup}). The fluctuations of the transmitted light are detected with a fast photodiode and analyzed with a narrow resolution bandwidth spectrum analyzer. The wavelength of the light is chosen so as to maximize the sensitivity of the interferometer to the displacement of the membrane modes considered~\cite{Nair2017}. The mechanical resonance frequencies can be determined by Lorentzian fits to the thermal noise spectrum, typically recorded with a resolution bandwidth of 0.5 Hz and averaged 500 times. The mechanical quality factors are determined either from the results of the Lorentzian fits to the thermal noise spectrum, or by performing ringdown spectroscopy of the resonantly excited mode~\cite{Nair2017,Naesby2017}.

\section{Mechanical frequency tuning}

\begin{figure}[h!]
\centering\includegraphics[width=\columnwidth]{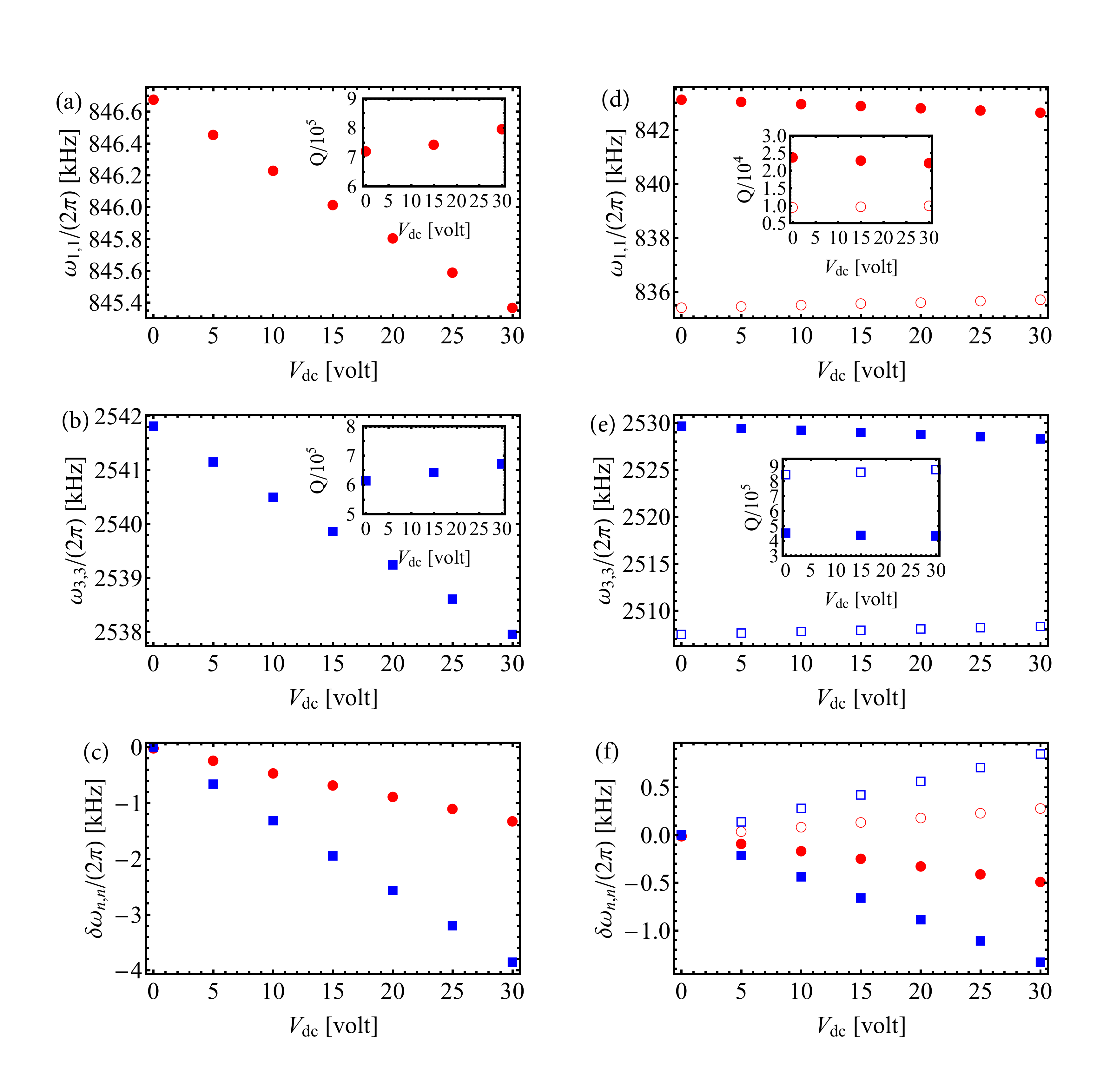}
\caption{(a,b) Measured \textit{single membrane} (1,1) and (3,3) mode frequencies as a function of bias voltage $V_{dc}$. (c) Corresponding resonance frequencies shifts versus $V_{dc}$. (d,e) Measured (1,1) and (3,3) mode resonance frequencies of the A (red circles and empty blue squares) and B (red dots and filled blue squares) membranes in a \textit{double-membrane} array as in Fig.~\ref{fig:scheme}c, as a function of $V_{dc}$. (f) Corresponding resonance frequency shifts versus $V_{dc}$. The error bars are smaller than the size of the data points. The insets in (a,b) and (d,e) show the variations of the mechanical quality factors with $V_{dc}$.}
\label{fig:actuationsingle}
\end{figure}

The application of a positive voltage to a single membrane chip with a 200 $\mu$m-thick frame, as depicted in Fig.~\ref{fig:scheme}b, results in a reduction of the tensile stress of the SiN film. In this tensile stress-dominated regime the resonance (angular) frequencies of the mechanical modes are to a very good approximation given by 
\begin{equation}\omega_{m,n}=\sqrt{\frac{\mathcal{T}}{\rho}}\frac{\pi}{a}\sqrt{m^2+n^2},\end{equation} where $\mathcal{T}$ is the tensile stress, $\rho$ the density of SiN, $a$ the membrane lateral dimension and $m$, $n$ strictly positive integers, and can thus be tuned by the application of a bias voltage $V_{dc}$ to the piezeoelectric element, as demonstrated in~\cite{Wu2018}. A linear reduction with $V_{dc}$ of the resonance frequencies of both the fundamental ($m=1,n=1$) and a higher order ($m=3,n=3$) modes is observed (Figs.~\ref{fig:actuationsingle}a and b), with frequency shifts of -43 Hz/V and -130 Hz/V, respectively. Furthermore, as shown in Fig.~\ref{fig:actuationsingle}c, the mechanical quality factors are observed to be only weakly (slight increase for this sample) dependent on the bias voltage.

We now turn to the piezoelectric actuation of double-membrane arrays as depicted in Figs.~\ref{fig:scheme} c and d. The compression of the PZT affects the whole Si frame/SiN films structure. In both geometries the lower (closer to the piezoelectric element) membrane sees its tensile stress increase when a positive dc-voltage is applied to the PZT, whereas the upper (further away from the piezoelectric element) sees its stress decrease. Figures~\ref{fig:actuationsingle}d and e show the variations with the bias voltage of the (1,1) and (3,3) modes of an array with a 200 $\mu$m-thick frame. Shifts of -16 Hz/V and -44 Hz/V are respectively observed for the (1,1) and (3,3) modes of the lower membrane, and 9 Hz/V and 29 Hz/V for those of the upper membrane. The mechanical quality factors are observed to be essentially independent of the bias voltage (Fig.~\ref{fig:actuationsingle}f).

\begin{figure}[h]
\centering\includegraphics[width=\columnwidth]{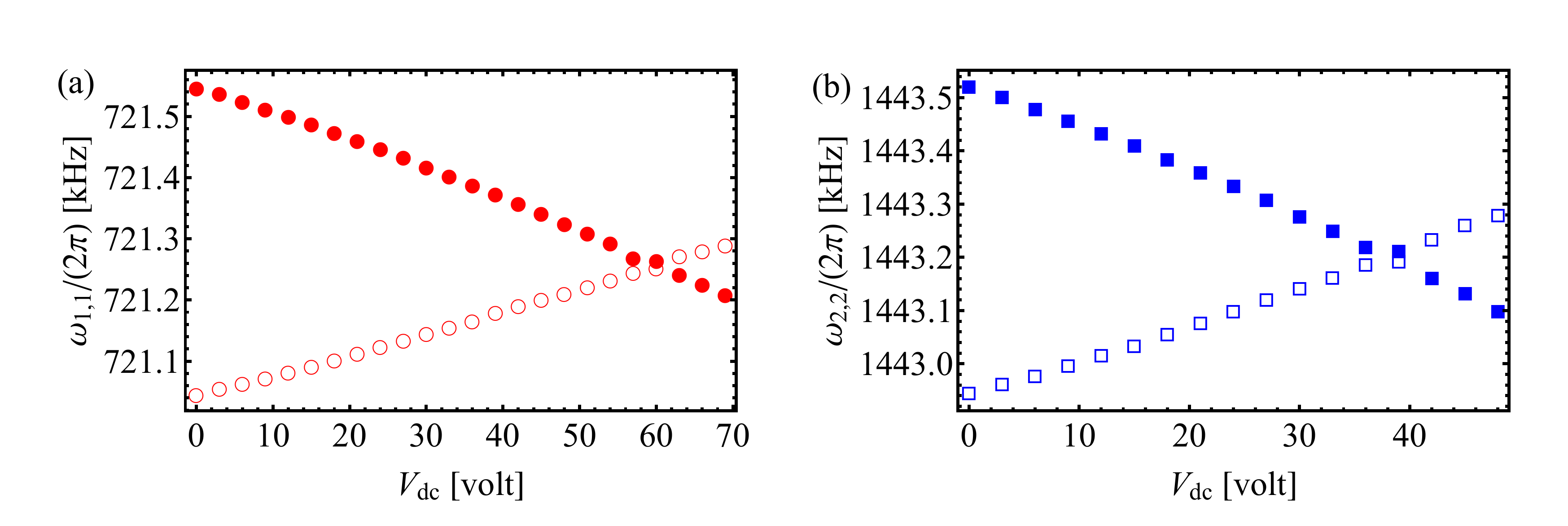}
\caption{(a) Measured (1,1) mode resonance frequencies of the A (red circles) and B (red dots) membranes in a double-membrane array as in Fig.~\ref{fig:scheme}d, as a function of $V_{dc}$. (b) Measured (2,2) mode resonance frequencies of the A (empty blue squares) and B (filled blue squares) membranes, as a function of $V_{dc}$.}
\label{fig:actuationmedium}
\end{figure}

Piezoelectric tuning of the modes of a shorter ($\sim 8.5$ $\mu$m) array with thicker frame (500 $\mu$m) and in the geometry of Fig.~\ref{fig:scheme}d was also demonstrated and reported in Ref.~\cite{Naserbakht2019}. Figures~\ref{fig:actuationmedium} shows the variations with $V_{dc}$ of the frequencies of the (1,1) and (2,2) modes of both membranes. The two membranes of this array exhibit fairly similar bare mechanical properties; for instance, the fundamental modes of the upper (A) and lower (B) membranes have resonance frequencies $721.05$ kHz and $721.55$ kHz, respectively, in absence of biasing. It is then possible to tune them to degeneracy for a bias voltage of approximately 56 V, as shown in Fig.~\ref{fig:actuationmedium}. The mechanical quality factors are also observed to be essentially independent of the bias voltage for this sample~\cite{Naserbakht2019}. A careful analysis of the evolution of the thermal noise spectra allows furthermore for extracting the intermembrane coupling, as discussed in Ref.~\cite{Naserbakht2019}. 

Having demonstrated the possibility to noninvasively tune the mechanical mode spectrum of such vertically coupled membrane systems, we now turn to their dynamical activation under parametric excitation.

\section{Parametric amplification of the thermal fluctuations}

\subsection{Theoretical model}

We consider a parametric modulation of the spring constant of a normal mode of the form $k=k_0[1+\xi\sin(\omega_pt)]$, where $k_0$ is the spring constant in absence of modulation, $\xi$ the normalized modulation amplitude, which is proportional to the applied modulation voltage amplitude and $\omega_p$ the pump modulation frequency. The classical dynamics of the amplitude around equilibrium, $x(t)$, can be modelled by the following differential equation~\cite{Rugar1991}
\begin{equation}
\ddot{x}+\Gamma\dot{x}+\omega_m^2[1+\xi\sin(\omega_p t)]x=F_{\textrm{th}}/m,
\label{eq:x}
\end{equation}
where $\omega_m=\sqrt{k_0/m}$ is the bare mechanical resonance frequency of the normal mode considered, $m$ the effective mass of the mode, $\Gamma=\omega_m/Q$ its damping rate and $F_\textrm{th}$ a thermal noise force arising from the coupling with the thermal environment. For a high quality factor oscillator ($\omega_m\gg\Gamma$) and a pump modulation frequency $\omega_p=2(\omega_m+\delta)$ close to the second harmonic frequency, it is convenient to introduce the slowly varying envelope $A(t)$ defined by
\begin{align}
x(t)&=\left[A(t)e^{-i\omega_mt}+A^*(t)e^{i\omega_mt}\right]/2\\
&=X_1(t)\cos(\omega_mt)+X_2(t)\sin(\omega_mt)
\end{align}
where the quadratures $X_1$ and $X_2$ are respectively given by the real and imaginary parts of $A$. Under the rotating wave approximation, Eq.~(\ref{eq:x}) yields
\begin{equation}
\dot{A}=-\left(\frac{\Gamma}{2}+i\delta\right)A+\omega_m\frac{\xi}{4}A^*+\tilde{F},
\label{eq:A}
\end{equation}
where $\tilde{F}=F_\textrm{th}e^{i\omega_mt}/m\omega_m$. Fourier transforming Eq.~(\ref{eq:A}) using the convention $f(\omega)=\int f(t)e^{-i\omega t}dt$ yields
\begin{equation}
\left(\gamma+i\omega+i\delta\right)A(\omega)-\gamma\epsilon A^*(\omega)=\tilde{F}(\omega),
\label{eq:AFT}
\end{equation}
where $\gamma=\Gamma/2$ and $\epsilon=\omega_m\xi/4\gamma=Q\xi/2$. From (\ref{eq:AFT}) and its complex conjugate one readily obtains
\begin{align}
A(\omega)&=\frac{(\gamma+i\omega-i\delta)\tilde{F}(\omega)+\gamma\epsilon\tilde{F}^*(\omega)}{(\gamma+i\omega+i\delta)(\gamma+i\omega-i\delta)-(\gamma\epsilon)^2},\\
A^*(\omega)&=\frac{(\gamma+i\omega+i\delta)\tilde{F}^*(\omega)+\gamma\epsilon\tilde{F}(\omega)}{(\gamma+i\omega+i\delta)(\gamma+i\omega-i\delta)-(\gamma\epsilon)^2}.
\end{align}
The noise spectrum, $S_{X_\theta}(\omega)=\langle X_\theta (\omega) X_\theta (-\omega)\rangle$, of an arbitrary quadrature $X_\theta=X_1\cos(\theta)+X_2\sin(\theta)$ can be computed using the fact that \begin{equation}
\langle F_{\textrm{th}}(\omega)F_{\textrm{th}}(\omega')\rangle=4k_BTm\Gamma\delta(\omega+\omega'),\end{equation} such that only $\langle \tilde{F}(\omega)\tilde{F}^*(-\omega)\rangle$ and $\langle \tilde{F}^*(\omega)\tilde{F}(-\omega)\rangle$ are nonzero and both equal to $4k_BT\gamma/m\omega_m^2$, where $k_B$ is the Boltzmann constant and $T$ the thermal bath temperature. 

If, like in the experiment, one does not keep track of the quadrature angle $\theta$, one observes an average noise spectrum
\begin{equation}
\bar{S}(\omega)\equiv\bar{S}_{X_\theta}(\omega)=\left\langle A(\omega)A^*(-\omega)+A^*(\omega)A(-\omega)\right\rangle/4
\end{equation}
given by
\begin{equation}
\bar{S}(\omega)=\frac{4k_BT\gamma}{m\omega_m^2}\frac{\gamma^2(1+\epsilon^2)+\omega^2+\delta^2}{[\gamma^2(1-\epsilon^2)+\delta^2-\omega^2]^2+4\gamma^2\omega^2}.\label{eq:Sav}
\end{equation}
At the parametric resonance ($\delta=0$), the average noise spectrum (\ref{eq:Sav}) can be simply written as 
\begin{equation}
\bar{S}_{\delta=0}(\omega)=\frac{2k_BT\gamma}{m\omega_m^2}\left[\frac{1}{\gamma^2(1-\epsilon)^2+\omega^2}+\frac{1}{\gamma^2(1+\epsilon)^2+\omega^2}\right]
\end{equation}
i.e. the sum of two Lorentzians with HWHMs $\gamma(1-\epsilon)$ and $\gamma(1+\epsilon)$, corresponding to the noise spectrum of the amplified and deamplified quadratures, respectively. While in general not Lorentzian the average noise spectrum is to a good approximation Lorentzian either at low gains ($\epsilon\ll 1$) or close to the parametric threshold $\epsilon\rightarrow 1$, where the contribution of the amplified quadrature noise dominates. At the parametric threshold $\epsilon=1$--or, equivalently, $\xi=2/Q$--the linewidth of the noise spectrum diverges (in absence of nonlinearities).

The variance of the amplitude, $\Delta x^2=\langle x^2\rangle$, is obtained by integrating this one-sided noise spectrum over the positive frequencies 
\begin{equation}
\Delta x^2=\int_0^{\infty} \bar{S}(\omega)\frac{d\omega}{2\pi}
\end{equation}
to give the average energy in the mode $\bar{E}=\frac{1}{2}m \omega_m^2 \langle x^2\rangle$ as
\begin{equation}
\bar{E}=\frac{\bar{E}_0}{1-\epsilon^2/(1+\delta^2/\gamma^2)},
\label{eq:var}
\end{equation}
where $\bar{E}_0=\frac{1}{2}k_BT$ is the average energy in absence of parametric modulation. In presence of parametric modulation the average energy thus increases with the modulation amplitude as $1/(1-\epsilon^2)$ and diverges at the threshold. For a nonzero pump detuning, the amplification is reduced with respect to that on resonance according to Eq.~(\ref{eq:var}) and the parametric resonance linewidth depends on both $\Gamma$ and $\epsilon$.

\subsection{Experimental results}

\begin{figure}[h!]
\centering\includegraphics[width=0.5\columnwidth]{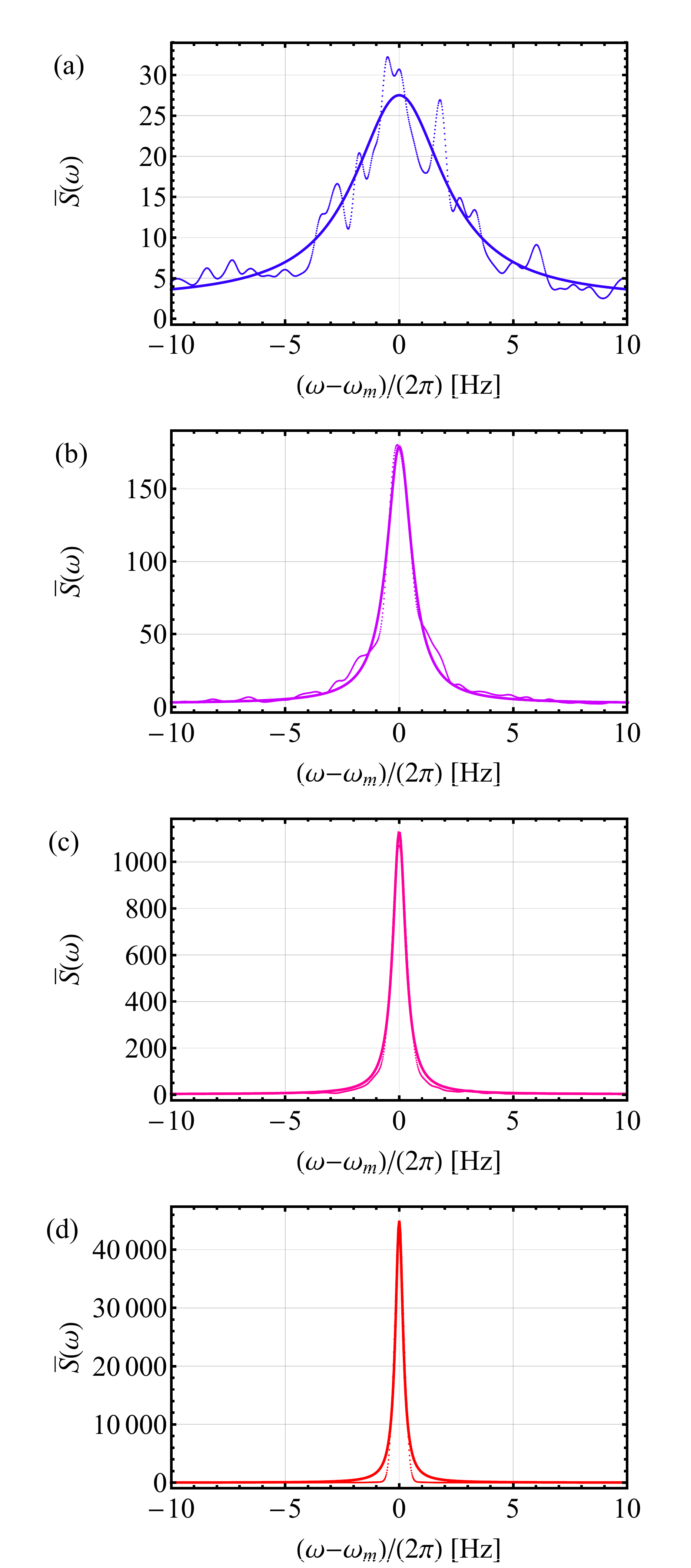}
\caption{Average noise spectra of membrane B's fundamental mode at a bias voltage corresponding to $\Delta/(2\pi)=30$ Hz and for different resonant modulation amplitudes corresponding to (a) $\epsilon=0.069$, (b) $\epsilon=0.809$, (c) $\epsilon=0.908$, (d) $\epsilon=0.996$. The solid lines are the results of Lorentzian fits to the data. The $y$-axis scale, in arbitrary units, is the same for all figures. The same RBW of 0.5 Hz is used for recording all the spectra.}
\label{fig:spectra}
\end{figure}

Parametric excitation of the membranes of the $8.5$ $\mu$m-long array, whose dc-actuation was shown in Fig.~\ref{fig:actuationmedium}, was investigated for different values of the biased voltage by applying a resonant parametric modulation at the second harmonic frequency of the fundamental mode of one of the membranes. The evolution of the noise spectrum of the fundamental modes was then monitored as a function of the modulation voltage amplitude $V_{ac}$. The bias voltage $V_{dc}$, which determined the frequency separation between the fundamental mode frequencies $\Delta=\omega_{1,1}^A-\omega_{1,1}^B$ was chosen in such a way that the second harmonic of the excited fundamental mode frequency, $2\omega_{1,1}^A$ or $2\omega_{1,1}^B$, did not coincide with a (2,2) mode resonance frequency. We verified experimentally that neither the non-resonantly driven (1,1) mode nor the (2,2) modes were not excited in presence of the parametric drive.

Figure~\ref{fig:spectra} shows examples of noise spectra of membrane B's fundamental mode, obtained for a bias voltage corresponding to a frequency separation $\Delta/(2\pi)=30$ Hz, and for increasing modulation amplitudes. The modulation amplitudes were normalized to the parametric threshold voltage and the same resolution bandwidth of 0.5 Hz is used for all spectra. The results of Lorentzian fits to the data are also shown. It can be seen that the amplified noise spectra are generally well-approximated by Lorentzians with increasing peak value/area and reduced linewidth, as the parametric modulation amplitude is increased. 

\begin{figure}[h]
\centering\includegraphics[width=\columnwidth]{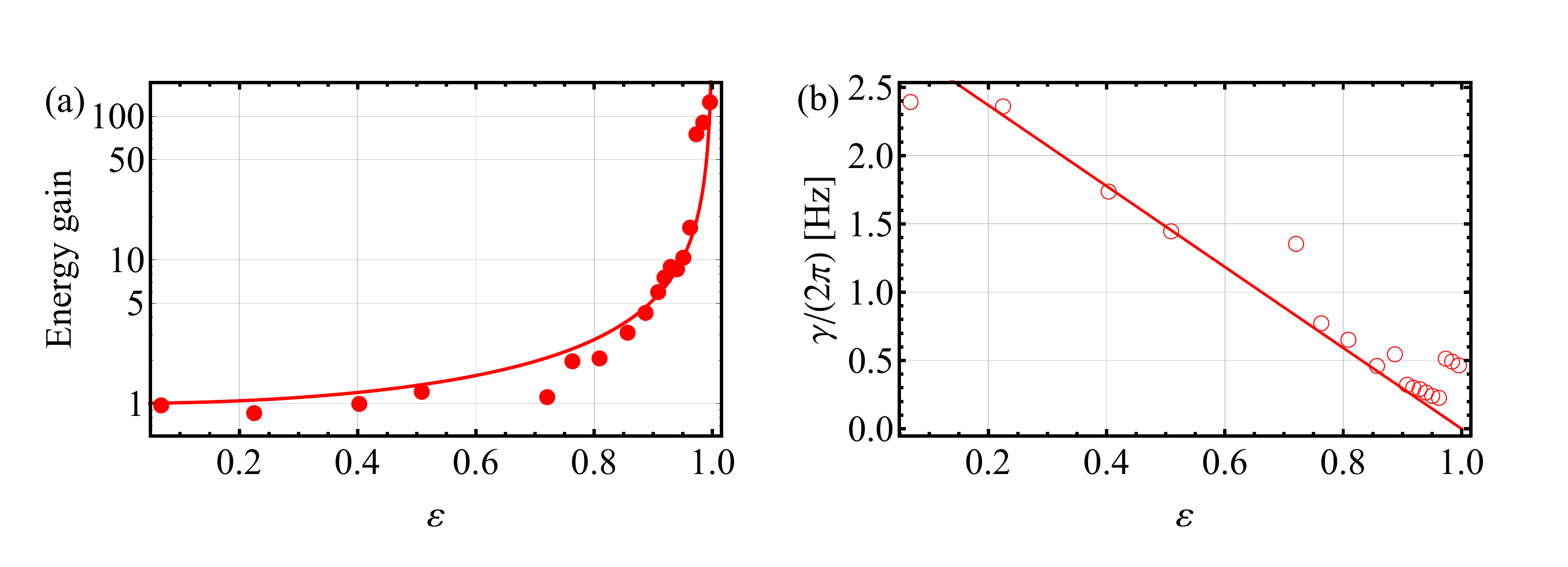}
\caption{(a) Energy gain $\bar{E}/\bar{E}_0$ as a function of the normalized parametric modulation amplitude $\epsilon$ for membrane B's fundamental mode at a bias voltage corresponding to $\Delta/(2\pi)=30$ Hz. The solid line shows the result of a fit according to Eq.~(\ref{eq:var}). (b) HWHM of the noise spectra resulting from the Lorentzian fits, as a function of $\epsilon$. The solid line shows a linearly decreasing function of the form $\gamma_B(1-\epsilon)$.}
\label{fig:gainwidth}
\end{figure}

The average energy is obtained by numerical integration of the spectrum and the energy gain--i.e. the ratio $\bar{E}/\bar{E}_0$ of the average energy for a given $\epsilon$ to the average energy for $\epsilon=0$--is shown in Fig.~\ref{fig:gainwidth}a as a function of $\epsilon$. The variations with $\epsilon$ of the HWHM obtained from the Lorentzian fits is also shown in Fig.~\ref{fig:gainwidth}b. The solid line in Fig.~\ref{fig:gainwidth}a shows the result of a fit according to Eq.~(\ref{eq:var}), while that in Fig.~\ref{fig:gainwidth}b shows a linearly decreasing linewidth of the form $\gamma_B(1-\epsilon)$, where $\gamma_B$ is the HWHM in absence of modulation. As one approaches the threshold, effects due to the finite resolution bandwidth of the spectrum analyzer become visible and the measured linewidth is effectively limited by the RBW. Note that this does not affect the determination of the average energy, however, since the latter is related to the numerical integration of the noise spectrum and is independent of the RBW. Gains in energy of about 100 were observed for this mode at this particular bias voltage, before the parametric oscillation threshold was reached. Similar behaviors--albeit with different gains and thresholds, as will be discussed further--were observed for both fundamental modes and different bias voltages.

\begin{figure}[h]
\centering\includegraphics[width=0.5\columnwidth]{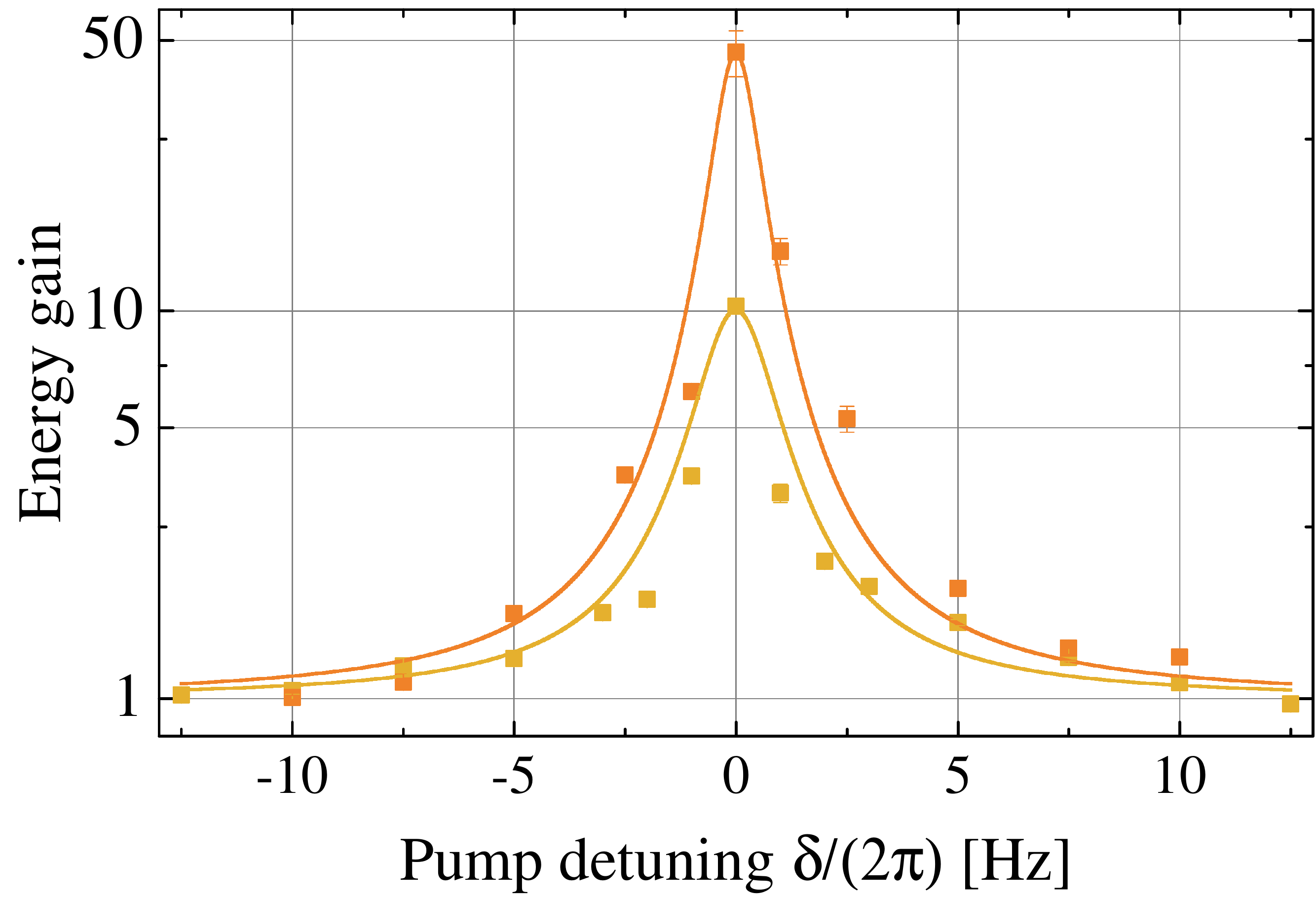}
\caption{Energy gain as a function of pump detuning $\delta$ for the fundamental mode of membrane B for $\Delta/(2\pi)=250$ Hz and for two modulation voltage amplitudes corresponding to $\epsilon=0.949$ (yellow) and $\epsilon=0.989$ (orange). The solid lines are the theoretical predictions of Eq.~(\ref{eq:var}).}
\label{fig:resonance}
\end{figure}

The resonant nature of the parametric excitation was also verified by scanning the modulation voltage frequency around the second harmonic frequency. Figure~\ref{fig:resonance} shows the variation of the energy gain--deduced from the noise spectra as previously--as a function of the pump detuning $\delta$, for the fundamental mode of membrane B at a bias voltage such that $\Delta/(2\pi)=250$ Hz and for two different parametric modulation amplitudes close to the threshold. The solid lines show the theoretical predictions of Eq.~(\ref{eq:var}), in which $\epsilon$ is determined by the zero-detuning value and $\gamma$ is the HWHM measured in absence of modulation.

\begin{figure}[h]
\centering\includegraphics[width=0.55\columnwidth]{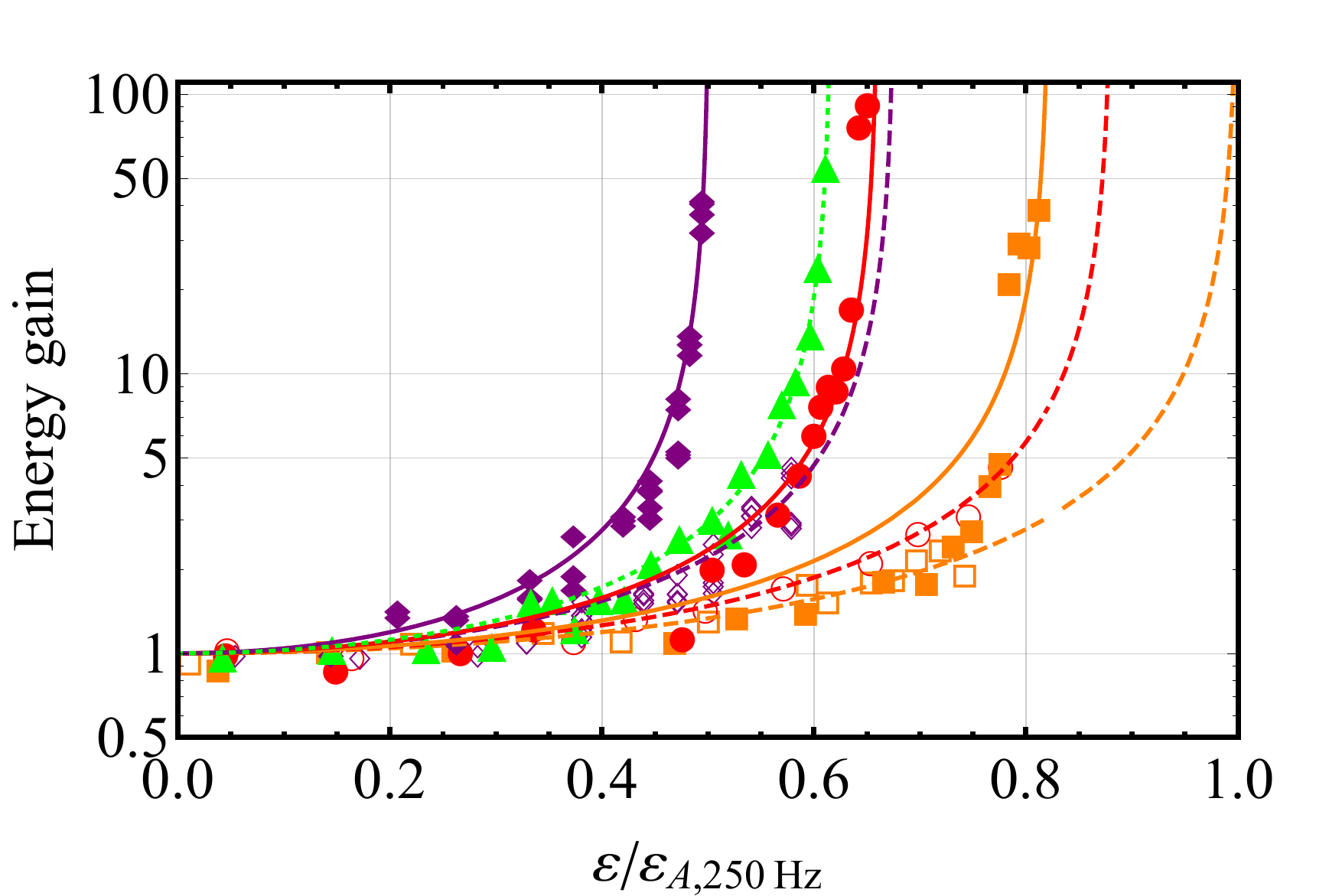}
\caption{Energy gains as a function of normalized modulation amplitude for both fundamental modes for increasing $V_{dc}$, such that $\Delta/(2\pi)=250$ Hz (yellow squares), 30 Hz (red circles) and -100 Hz (purple diamonds). The empty and full symbols are for membranes A and B, respectively. The green triangles correspond to the energy gain measured at the degeneracy ($\Delta=0$). The modulation amplitude is arbitrarily normalized to the threshold amplitude for membrane A's fundamental mode at $\Delta/(2\pi)=250$ Hz. The solid lines show the results of fits to Eq.~(\ref{eq:var}).}
\label{fig:summary}
\end{figure}

We finally investigated the influence of the bias voltage $V_{dc}$ on the resonant parametric amplification of the fundamental modes of both membranes. The energy gains of both modes for different bias voltages are shown in Fig.~\ref{fig:summary} as a function of the resonant modulation amplitude. The amplitude was arbitrarily normalized to the parametric amplification threshold amplitude, $V_{ac}=448$ mV, of membrane A's fundamental mode at $\Delta/(2\pi)=250$ Hz. One first observes that increasing the bias voltage leads to lower parametric oscillation thresholds for both membranes. Higher thresholds for membrane A than for membrane B at comparable $V_{dc}$s were also systematically observed. The reasons for this behavior were investigated in detail in Ref.~\cite{Naserbakht2019}. In brief, the lower membrane (B) experiences a direct modulation of its tensile stress at the parametric resonance frequency through the compression of the lower frame. The parametric oscillation threshold can be reached by the application of a moderate modulation voltage. In contrast, such a direct modulation of the SiN film spring constant is much weaker for the upper membrane (A). However, the application of a static compressive force to the chips results in a modification of the nonlinear response/stress of membrane A, which accounts for the lowering of the parametric amplification threshold with the bias voltage. Let us also note that, due to thermal drifts during the measurements, variations in the highest achievable gains close to the threshold were typically observed, so that one cannot rule out that higher subthreshold gain values may be achievable by better temperature control. Let us also remark that, at degeneracy ($\Delta=0$), the single parametrically excited mode picture breaks down, as the dynamics of the two coupled modes have in principle to be taken into account~\cite{Mahboob2014}. However, since the parametric gain for membrane B is substantially stronger than for membrane A, the total amplified noise spectrum becomes quickly dominated by that of membrane B's fundamental mode, as the modulation amplitude increases. The observed parametric gains and threshold values thus consistently follow those of membrane B.

\section{Conclusion}

Noninvasive tuning of the mechanical resonance frequencies of suspended square drum resonators in monolithic vertical arrays was demonstrated using a simple scheme where the membrane chips were directly mounted on a piezoelectric ring actuator. The application of a piezoelectrically controlled force to the bottom chip allowed for modifying the tensile stress of the membrane resonators and thereby change their frequencies. For membranes with not too different bare mechanical frequencies, tuning to degeneracy by the application of a dc-voltage to the PZT is possible, as the bottom and top membrane experience opposite frequency shifts with the bias voltage. 

Dynamical actuation of both membranes was also demonstrated by the application of an ac-voltage at twice the mechanical resonance frequencies of the fundamental modes of the membranes and observing the parametric amplification of their thermal fluctuations until the parametric oscillation threshold was reached. The experimental observations are well-accounted for by a simple phase-averaged subthreshold parametric oscillator model. Last, the amplification was shown to be enhanced by the simultaneous application of the dc-voltage, which results in higher amplification gains and a lowering of the parametric oscillation thresholds.

Such a tuning of both the linear and nonlinear response of membrane resonators in vertical arrays is promising for exploring collective effects and investigating phonon dynamics in currently investigated optomechanical arrays~\cite{Nair2017,Piergentili2018,Gartner2018,Wei2019}, or for sensing applications~\cite{Naesby2017,Naesby2018,Naserbakht2019sna}.

\begin{acknowledgements}
We thank the Velux Foundations for support.
\end{acknowledgements}

\bibliographystyle{apsrev}
\bibliography{parametric_bib}

\end{document}